\documentstyle[12pt]{article}
\newcommand{\be}{\begin{equation}}
\newcommand{\ee}{\end{equation}}
\newcommand{\bea}{\begin{eqnarray}}
\newcommand{\eea}{\end{eqnarray}}
\newcommand{\ba}{\begin{array}}
\newcommand{\ea}{\end{array}}

\def\bbox{{\,\lower0.9pt\vbox{\hrule \hbox{\vrule height 0.2 cm
\hskip 0.2 cm \vrule height 0.2 cm}\hrule}\,}}
\newcommand{\dsl}{\pa \kern-0.5em /}

\newcommand{\nn}{\nonumber \\}
\newcommand{\p}[1]{(\ref{#1})}

\font\mybb=msbm10 at 10pt
\def\bb#1{\hbox{\mybb#1}}

\def\bE {\bb{E}}

\begin{document}

%%%%%%%%%%%%%%%% title page %%%%%%%%%%%%%%%%%%%%%%%%%%%%%%%%%%%%

%%%%%%%%%%%%%%%% title page %%%%%%%%%%%%%%%%%%%%%%%%%%%%%%%%%%%%

\begin{titlepage}
\rightline{UB-ECM-PF-04/08}
\rightline{\tt{hep-th/0404108}}
%\begin{flushright}
%\end{flushright}

\vfill

\begin{center}
\baselineskip=16pt
{\Large\bf  A Super-Flag Landau Model{$^\star$}}
\vskip 0.3cm
{\large {\sl }}
\vskip 10.mm
{\bf ~Evgeny Ivanov$^{*}$, Luca Mezincescu$^{\dagger}$ and  Paul K.
Townsend$^{+,\natural}$}
\vskip 1cm
{\small
$^*$
Bogoliubov Laboratory of Theoretical Physics \\
JINR, 141980 Dubna, Russia\\
}
\vspace{10pt}
{\small
$^\dagger$
Department of Physics,\\
University of Miami,\\
Coral Gables, FL 33124, USA\\
}
\vspace{10pt}
{\small
  $^+$
Instituci\'o Catalana de Recerca i Estudis Avan\c cats, \\
Departament ECM, Facultat de F\'{\i}sica, \\
Universitat de Barcelona, Diagonal 647,\\
E-08028 Barcelona, Spain
}
\end{center}
\vfill

\par
\begin{center}
{\bf ABSTRACT}
\end{center}
\begin{quote}

We consider the quantum mechanics of a particle on the coset superspace
$SU(2|1)/[U(1)\times U(1)]$, which is a super-flag manifold with
$SU(2)/U(1)\cong S^2$ `body'. By incorporating the Wess-Zumino terms associated with the
$U(1)\times U(1)$ stability group, we obtain an exactly solvable
super-generalization  of the Landau model for a charged particle on the
sphere. We solve this model using the factorization method. Remarkably,
the physical Hilbert space is finite-dimensional because the number of admissible Landau levels
is bounded by a combination of the $U(1)$ charges. The level saturating the
bound has a wavefunction in a shortened, degenerate, irrep of $SU(2|1)$.

\vfill
\vfill
\vfill

$^\star$ {To appear in Ian Kogan Memorial Volume,
{\it From Fields to Strings: Circumnavigating Theoretical Physics},
eds. M.
Shifman, A. Vainshtein, and J. Wheater, World Scientific (2004) }

\vfill
  \hrule width 5.cm
\vskip 2.mm
{\small
%\noindent $^1$ eivanov@thsun1.jinr.ru\\
%\noindent $^2$ mezincescu@server.physics.miami.edu\\
\noindent $\natural$ On leave from Department of Applied
Mathematics and Theoretical Physics, University of Cambridge, UK.
\\ }
\end{quote}
\end{titlepage}
%%%%%%%%%%%%%%%%%%%%%%%%%%%%%%%%%%%%%%

\setcounter{equation}{0}
\section{Introduction}
\label{sec.intro}

In 1930 Landau posed and solved the problem of a quantum particle in a plane orthogonal
to a uniform magnetic field, showing in particular that the particle's energy is restricted
to a series of `Landau levels'
\cite{landau}. It is now customary to call a `Landau model' any problem in which
a quantum particle is confined to a surface orthogonal to a magnetic field that is uniform on the surface.
A case in point is the
Landau model of a particle on a unit sphere in $\bE^3$ with a magnetic monopole at the centre.
This model was introduced by Haldane in the context of the Quantum Hall Effect \cite{haldane},
and has many fascinating features. For example, it is exactly soluble \cite{stone}. When restricted
to the lowest Landau level (LLL) the sphere becomes the phase space rather than the configuration space,
and this leads to a physical realization of the fuzzy sphere \cite{madore}.

Ian Kogan worked on aspects of Landau models \cite{kogan1} around the same time that he developed
the idea of the `monopole bag' \cite{kogan3} in which a  closed axion domain wall is supported against
collapse by the electric charge induced on it by a magnetic monopole inside. Perhaps he saw a connection?
The `monopole bag' was what inspired  one of us to observe that a  closed D2-brane carrying
a net electric charge would appear to be a D0-brane \cite{pkt}, and it is now appreciated that
there are circumstances in which it is energetically favourable for D0-branes to `expand'
into a fuzzy spherical D2-brane \cite{myers,HNS}. The fuzzy sphere thus appears as a common theme.

Recently, we showed how the fuzzy {\it supersphere} emerges from the LLL quantum mechanics
of a  particle on the coset superspace $SU(2|1)/U(1|1)$ \cite{IMT}. There is a natural
extension  of this model to a full Landau model but this involves terms quadratic in time-derivatives
of the Grassmann odd variables, and such terms would normally be considered `higher-derivative'.
This is one of the reasons that supergroups such as $SU(2|1)$ do not normally appear as symmetry
groups in physical problems.

Here we show that `higher-derivative' fermion terms {\it can} be avoided in an $SU(2|1)$-invariant
extension of the full Landau problem for a particle on the sphere,
but instead of the supersphere one has to consider the coset superspace
\be
SU(2|1)/[U(1)\times U(1)] \equiv SF.\label{Main}
\ee
This again has $SU(2)/U(1)\cong S^2$ `body' and is a homogeneous K\"ahler superspace, but
it is not a symmetric superspace. It is a flag supermanifold, analogous
to the flag manifold $SU(3)/[U(1)\times U(1)]$. For the sake of brevity, we call it the `super-flag' (SF).
This super-extension of the sphere allows the construction of a Landau-type model with a `canonical'
fermion kinetic term arising from Wess-Zumino (WZ) terms associated with the two $U(1)$ factors of
the stability subgroup. The phase space of this model has real dimension $(4|4)$,
so the configuration space has real dimension $(2|2)$ with $S^2$ body, exactly as one would have
for a particle on the supersphere, but without the `higher-derivative' fermion kinetic term.

We quantize this model using techniques explained recently in \cite{IMPT, IMT}:
this leads to a Hilbert space spanned by `chiral'  superfields on SF. The Hamiltonian
is shown to act in this physical subspace and we use Schroedinger's factorization method \cite{Sch}
to determine its eigenstates and
eigenvalues, following the application of this method to the Landau model for a particle
on the sphere \cite{MM}. Remarkably, we find that the number of Landau levels is finite,
in contrast to the infinite number of levels in the bosonic case. This is because wavefunctions
with positive norm
exist only for $\ell \leq 2M$, where $\ell$ is the number of the Landau level and $M$
is the properly normalized positive eigenvalue of some combination of two $U(1)$ charges.
The full Hilbert space is therefore finite dimensional!

\setcounter{equation}{0}
\section{Super-flag geometry}
\label{sec.geometry}

The supergroup $SU(2|1)$ can be defined as the group of $(1|2)\times (1|2)$
unitary supermatrices of unit super-determinant. A parametrization of $SU(2|1)$
that makes manifest the K\"ahler property of its coset superspace $SU(2|1)/[U(1)\times U(1)]$
can be found following steps analogous to those spelled out for $SU(3)/[U(1)\times U(1)]$  in \cite{picken}.
The group
$SU(2|1)$ acts linearly on vectors in a vector superspace of dimension $(1|2)$.
A simple choice of basis in this superspace is provided by the columns of the supermatrix
\bea
\left( \begin{array}{ccc}
           1   &   0  &  0\\
    -\xi^2   &   1  &  0\\
    -\xi^1   &   z   &  1
    \end {array}
    \right )
  \eea
where $z$ is a complex variable and $\xi^i$ ($i=1,2$)
are complex anticommuting variables, with complex conjugates $\bar\xi_i$.
By an application of the Gramm-Schmidt procedure we can transform the above supermatrix
into a unitary supermatrix $U$ for which the three column supervectors are orthonormal.
This ensures that $U \in SU(2|1)$. One finds that
\be\label{Umatrix}
U= \pmatrix{ {1\over {K_{1}}^{1\over 2}}\left[
\begin{array}{c}
           1   \\
    -\xi^2  \\
    -\xi^1
    \end {array}
    \right ] &
{\left({K_{1}\over {K_{2}}}\right)^{1\over 2}}\left[
\begin{array}{c}
   \left({{\bar \xi}_{2} + z {\bar \xi}_{1}}\right)/ {{K_{1}}^{2}}  \\
    1 - {\bar \xi}_{1}\left({\xi^{1}} - {z\xi^{2}} \right)  \\
      z + {\bar \xi}_{2}\left({\xi^{1}} - {z\xi^{2}} \right)
    \end {array}
    \right ]  &
{1\over {K_{2}}^{1\over 2}}\left[ \begin{array}{c}
       {\bar \xi}_{1} - {\bar z}  {\bar \xi}_{2}   \\
    -{\bar z} \\
    1
    \end {array}
    \right ]}
\ee
where
\be
K_{1} = 1 + {\bar \xi}_{1} \xi^{1} +  {\bar \xi}_{2} \xi^{2}, \qquad
K_{2} = 1 +{\bar z}z + \left({\xi^{1}} - {z\xi^{2}} \right)
\left( {\bar \xi}_{1} - {\bar z}  {\bar \xi}_{2}\right).
\ee
The general $SU(2|1)$ supermatrix can be written in the form $Uh$,
where $h$ is a diagonal unitary supermatrix with unit superdeterminant parametrized by two angles.
This means that the unitary supermatrix $U$ provides a parametrization of
the coset superspace $SU(2|1)/[U(1)\times U(1)]$.

To compute the Cartan forms and $U(1)$ connections for $SU(2|1)/[U(1)\times U(1)]$,
we write the Lie superalgebra valued 1-form
$U^{-1}dU$ as
\be
U^{-1}dU \equiv \Omega = \pmatrix{0 & \bar E_2 & \bar E_1\cr -E^2 & 0 &
-\bar E_+\cr -E^1 & E^+ & 0 } - \frac{i}{2}\pmatrix{{\cal B}&0&0 \cr 0& {\cal B}-{\cal A} & 0
\cr 0&0& {\cal A}}.
\ee
The Cartan 1-forms are $E^A=(E^+,E^1,E^2)$ and their complex conjugates are
$\bar E_A = (\bar E_+,\bar E_1,\bar E_2)$. One finds that
\be
E^A = dZ^ME_M{}^A, \qquad \bar E_A = d\bar Z_M \bar E^M{}_A,
\ee
where $Z^M=(z,\xi^1,\xi^2)$ are the complex coordinates and $\bar Z_M=
(\bar z, \bar\xi_1,\bar\xi_2)$ their complex conjugates; this defines the
(complex) supervielbein $E_M{}^A$. Using the inverse supervielbein
$E_A{}^M$, and its complex conjugate $\bar E^A{}_M$, we define the
complex supercovariant derivative $D_A$ and its complex conjugate $\bar D^A$ as
\be
D_A = E_A{}^M \partial_M, \qquad \bar D^A = \bar E^A{}_M \bar\partial^M.
\ee
A computation shows that
\bea\label{cartan}
E^+ &=& K_1^{-{1\over 2}} K_2^{-1}\left[dz - K_1^{-1}
\left( d\xi^1 - z d\xi^2\right)\left(\bar\xi_2 + z\bar\xi_1
\right)\right],
\nn
E^1 &=&  \left(K_1K_2\right)^{-{1\over2}}
\left[ d\xi^1 - z d\xi^2\right],
\nn
E^2 &=&  K_2^{-{1\over2}} \left[
  d\xi^1\left(\bar z -\xi^2\left(\bar\xi_1 - \bar z\bar \xi_2\right)
\right) +  d\xi^2\left(1 +\xi^1\left(\bar\xi_1
-\bar z\bar\xi_2\right)\right) \right]
\eea
and that
\bea\label{trans}
D_+ &=&  K_1^{1\over2}K_2\partial_z\,, \nn
D_1 &=& K_2^{1\over2}K_1^{-{1\over2}}\left(\bar\xi_2 +
z\bar\xi_1\right)\partial_z \nn
&& +\  K_1^{1\over2}K_2^{-{1\over2}}\left\{\left[1 + \xi^1
\left(\bar\xi_1 - \bar z \bar\xi_2\right)\right]\partial_{\xi^1} -
\left[\bar z - \xi^2\left(\bar\xi_1 - \bar z
\bar\xi_2\right)\right]\partial_{\xi^2}\right\}, \nn
D_2 &=&  K_{2}^{-{1\over2}}\left(z\partial_{\xi^1} +
\partial_{\xi^2}\right).
\eea
For the $U(1)$ connections ${\cal A}$ and ${\cal B}$ we have, similarly,
that
\be
{\cal A} = dZ^M {\cal A}_M + c.c. ,\qquad {\cal B} = dZ^M {\cal B}_M +
c.c.
\ee
and a calculation shows that
\be
{\cal A} = - i dZ^M\partial_M \log K_2 + c.c. , \qquad
{\cal B} = i dZ^M \partial_M \log K_1 + c.c.\,.
\ee

The $SU(2|1)$ transformations of the superspace coordinates $Z^M, \bar Z_M$ can be
found as follows. Let us write $U({\cal Z})$ for the unitary supermatrix
(\ref{Umatrix}) where
\be
{\cal Z} = (Z^M, \bar Z_M)\,.
\ee
For any element $U\in SU(2|1)$ we have
\be
UU({\cal Z}) = U({\cal Z}')h
\ee
for some diagonal unitary matrix $h$ in the $U(1)\times U(1)$ stability
subgroup. We choose $h$ to have the expansion 
\be
h = I + \left(\alpha \tilde{J}_3 + \beta \tilde{B}\right) + \cdots \label{hinfin}
\ee
where
\be
\tilde{J}_3 = \pmatrix{0 & 0 & 0\cr 0 & -1 &
0 \cr 0 & 0 & 1 }\,, \quad \tilde{B} = \pmatrix{1 & 0 & 0 \cr 0 & 1 &
0 \cr 0 & 0 & 0 }\,.
\ee
If one now chooses $U=U(\Delta, \bar\Delta)$ for constant infinitesimal parameter
$\Delta=(a,\epsilon^1,\epsilon^2)$, where $a$ is Grassmann-even and
$\epsilon^i$ ($i=1,2$) Grassmann-odd, then one finds that $Z' = Z+
\delta Z$, where
\bea\label{transf}
\delta  z &=& a + \bar a z^2 - \left(\bar\epsilon_2 +
z\bar\epsilon_1\right)\left(\xi^1 -z\xi^2\right), \nn
\delta \xi^1 &=& a \xi^2 + \epsilon^1 + \left(\bar\epsilon\cdot
\xi\right)\xi^1\,, \nn
  \delta \xi^2 &=& -\bar a\xi^1 + \epsilon^2 +
\left(\bar\epsilon\cdot\xi\right)
\xi^2
\eea
and
\bea
&& \alpha({\cal Z},\Delta, \bar\Delta)
= \frac{1}{2}\left[\bar a z - a\bar z +(\bar\xi_1 - \bar z \bar\xi_2)\epsilon^1
-\bar\epsilon_1(\xi^1 - z \xi^2) \right], \nn
&& \beta({\cal Z},\Delta, \bar\Delta) = \frac{1}{2}\left(\bar\xi\cdot\epsilon
- \bar\epsilon\cdot \xi\right)\,. \label{alphabeta}
\eea
The $U(1)\times U(1)$ transformations of the coordinates corresponding to ${\cal Z}$-independent
parameters $\alpha_0$ and $\beta_0$ in \p{hinfin} ($\bar\alpha_0 = -\alpha_0, \bar\beta_0 = -\beta_0$)
are as follows
\be
\delta z = (2\alpha_0 -\beta_0)\,z\,, \quad \delta \xi^1 =(\alpha_0 -\beta_0)\,\xi^1\,,\quad
\quad \delta \xi^2 = -\alpha_0\,\xi^2\,.\label{u1transf}
\ee
We have therefore shown that, in the chosen parametrization of the superflag,
the $SU(2|1)$ transformations of $(z,\bar z, \xi^i, \bar \xi_i)$ are analytic:
the coordinates $Z=(z, \xi^i)$ transform among themselves,
and the same is true for $\bar Z = (\bar z, \bar\xi_i)$.
Various other $SU(2|1)$ invariant subspaces determine the various types of superfields
that one can define on the superflag, as we now explain.

\setcounter{equation}{0}
\section{Super-flag superfields}
\label{sec.superfield}

In accord with the general procedure of nonlinear realizations, superfields
given on SF are characterized by two external $U(1)$ charges. The corresponding
operators ${\hat J}_3$ and $\hat B$ are the `matrix' parts
of the differential operators representing the $U(1)\times U(1)$ subgroup of $SU(2|1)$
(in other words, ${\hat J}_3$ and $\hat B$ count external $U(1)$ charges of the superfield).
The only superfields that we need to consider are those that are eigenfunctions of
$\hat J_3$ and $\hat B_3$ with eigenvalues $2N$ and $2M$, respectively:
\be\label{chargesf}
\hat{J}_3\, \Psi^{(N,M)}({\cal Z}) = 2N\,\Psi^{(N,M)}({\cal Z})\,, \quad
\hat{B}\,\Psi^{(N,M)}({\cal Z}) = 2M\,\Psi^{(N,M)}({\cal Z})\,.
\ee
Such superfields transform as
\be
\Psi^{(N,M)}{\;}'({\cal Z}') = h({\cal Z},\Delta, \bar\Delta) \Psi^{(N,M)}({\cal Z})\,.
\ee
In infinitesimal form,
\be\label{TranSf}
\delta \Psi^{(N,M)}({\cal Z}) = 2\left[N\, \alpha({\cal Z},\Delta,\bar\Delta)
+ M\,\beta({\cal Z},\Delta, \bar\Delta)\right]\Psi^{(N,M)}({\cal Z})\,.
\ee
%Let $\Psi$ be a general superfield on SF,
%and let

The $U(1)\times U(1)$ gauge covariant differential of a general superfield $\Psi$ on SF is
\be
{\cal D}\Psi = \left(d -\frac{i}{2}{\cal A}\hat{J}_3 -\frac{i}{2}{\cal B}\hat{B}\right)\Psi
= \left( E^A{\cal D}_A +\bar E_A\bar{\cal D}^A \right)\Psi,
\ee
which defines the gauge covariant derivatives ${\cal D}_A$. Using the identities
\bea
&& D_1K_2 = K_2^{\frac{3}{2}}K_1^{-\frac{1}{2}} \bar\xi_1\,, \quad
D_1K_1 = -K_1^{\frac{3}{2}}K_2^{-\frac{1}{2}}
(\bar\xi_1 -\bar z \bar\xi_2)\,, \nn
&& D_2 K_2 = 0\,, \qquad D_2K_1 = -K_2^{-\frac{1}{2}} (\bar\xi_2 + z \bar\xi_1)\,,
\eea
one finds that
\bea
&&{\cal D}_+ = D_+ - {1\over 2} K^{\frac{1}{2}}_1\partial_z K_2\,\hat{J}_3\,, \;\;
{\cal D}^+ = \overline{\cal D_+} = D^+ + {1\over 2}K^{\frac{1}{2}}_1\partial_{\bar z} K_2\,\hat{J}_3\,, \nn
&& {\cal D}_1 = D_1 - {1\over 2}K^{-\frac{1}{2}}_1K_2^{\frac{1}{2}}\bar\xi_1\,\hat{J}_3
- {1\over 2}K^{\frac{1}{2}}_1K_2^{-\frac{1}{2}}(\bar\xi_1 -\bar z \bar\xi_2)\,\hat{B}\,,\nn
&& \bar{\cal D}^1 = \bar D^1 + {1\over 2}K^{-\frac{1}{2}}_1K_2^{\frac{1}{2}}\xi^1\,\hat{J}_3
+ {1\over 2}K^{\frac{1}{2}}_1K_2^{-\frac{1}{2}}(\xi^1 -z \xi^2)\,\hat{B}\,, \nn
&&
{\cal D}_2 = D_2 - {1\over 2}K^{-1}_1K_2^{-\frac{1}{2}}(\bar\xi_2 + z \bar\xi_1)\,\hat{B}\,,\nn
&&
\bar{\cal D}^2 = \bar D^2 + {1\over 2}K^{-1}_1K_2^{-\frac{1}{2}}(\xi^2 +\bar z \xi^1)\,\hat{B}\,.
\eea
The geometry of the coset superspace $SU(2|1)/[U(1)\times U(1)]$ is now encoded in the
(anti)commutation relations
\bea
&& \left[{\cal D}_+, {\cal D}^+\right] = \hat{J}_3\,, \label{1A}\\
&& \left\{{\cal D}_1, {\cal D}_1 \right\} =
\left\{{\cal D}_2, {\cal D}_2 \right\}= \left\{{\cal D}_1, {\cal D}_2 \right\} = 0
\quad \mbox{and c.c.}\,, \label{1a} \\
&& \left\{{\cal D}_1, \bar{\cal D}^1 \right\} = (\hat{J}_3 + \hat{B})\,, \quad
\left\{{\cal D}_2, \bar{\cal D}^2 \right\} = \hat{B}\,, \label{1b} \\
&& \left\{{\cal D}_1, \bar{\cal D}^2 \right\} = -{\cal D}_+\,, \quad
\left\{{\cal D}_2, \bar{\cal D}^1 \right\} = {\cal D}^+\,, \label{1c}\\
&& \left[{\cal D}_+, \bar{\cal D}^1\right] = -\bar{\cal D}^2\,, \quad
\left[{\cal D}_+, \bar{\cal D}^2\right] = 0\,, \nn
&&
\left[{\cal D}^+, \bar{\cal D}^2\right] = \bar{\cal D}^1\,, \quad
\left[{\cal D}^+, \bar{\cal D}^1\right] = 0\,,  \label{1d}\\
&&
\left[{\cal D}_+, {\cal D}_1\right] = 0\,, \quad
\left[{\cal D}_+, \bar{\cal D}_2\right] = {\cal D}_1\,, \nn
&&
\left[{\cal D}^+, {\cal D}_2\right] = 0\,, \quad
\left[{\cal D}^+, {\cal D}^1\right] = -{\cal D}_2\,.\label{1e}
\eea

Using the fact that the charges of the covariant derivatives are opposite to those of the Cartan forms,
the $U(1)\times U(1)$ assignments of both can be worked out from the transformation rule
\be
\Omega{\;}' = h \Omega h^{-1} - d\alpha \tilde{J}_3 - d \beta \tilde{B}\,.
\ee
Here we record the result for the $U(1)$ charges of the covariant derivatives:
\bea
&& \hat{J}_3\,{\cal D}_+ = -2 {\cal D}_+\,, \;\;\hat{J}_3\,{\cal D}_1 = -{\cal D}_1\,, \;\;
\hat{J}_3\, {\cal D}_2 = {\cal D}_2\,, \label{ItransfD}\\
&& \hat{B}\, {\cal D}_+ = {\cal D}_+\,,\;\; \hat{B}\, {\cal D}_1 = {\cal D}_1\,,\;\;
\hat{B}\, {\cal D}_2 = 0\,.\label{BtransfD}
\eea
Note that, instead of $\hat{B}$, it is sometimes more convenient to use
the combination\footnote{This is just the matrix part of the $U(1)$ generator $J_3 + 2B$ that
commutes with the $SU(2)$ generators.}
\be
\hat{F} = 2\hat{B} +\hat{J}_3\,, \label{hatF}
\ee
which is distinguished by the fact that the $S^2$ covariant derivatives ${\cal D}_+,
{\cal D}^+(=\bar{\cal D}_+)$
(and the corresponding Cartan forms) have $\hat{F}$ charge zero, while
both spinor derivatives have  $\hat{F}$ charge 1:
\be
\hat{F}\, {\cal D}_2 = {\cal D}_2\,, \quad \hat{F}\, {\cal D}_1 = {\cal D}_1\,.
\ee

It will be convenient to set
\be
{\cal D}_+ = K_1^{\frac{1}{2}}K_2 \nabla^{(N)}_z\,, \quad
{\cal D}^+= K_1^{\frac{1}{2}}K_2 \nabla^{(N)}_{\bar{z}}\,,\label{Dnabla}
\ee
which defines the `semi-covariant' derivatives
\bea
&& \nabla_z^{(N)} = \partial_z -i N {\cal A}_z =
\partial_z - N \partial_z \log K_2\,, \nn
&& \nabla_{\bar z}^{(N)} = \partial_{\bar z} -i N {\cal A}_{\bar z} =
\partial_{\bar z} + N \partial_{\bar z} \log K_2\,. \label{semicov}
\eea
The $N$ dependence arises here because we assume that the covariant derivatives
act on superfields $\Psi^{(N,M)}$ obeying \p{chargesf}. It is easy to check that
\p{1A} is equivalent to the following commutation relation between the `semi-covariant' derivatives
\be\label{commnabla}
\left[\nabla_z^{(N)}, \nabla_{\bar{z}}^{(N)}\right] = 2K_1^{-1}K_2^{-2}\,N\,.
\ee
This can also be checked by using the identity
$$
\partial_z\partial_{\bar z}\log K_2 = K_1^{-1}K_2^{-2}\,.
$$

Let us now note a few important corollaries of the (anti)commutation relations:

\begin{itemize}

\item For any value of $N$ and $M$ it is consistent to consider covariantly chiral
or anti-chiral superfields\footnote{As an aside, let us note that, besides the chirality
conditions \p{CHIR}, one can consistently impose on a general $SU(2|1)$ superfield the
Grassmann analyticity conditions
${\cal D}_2\Psi = \bar{\cal D}^1\Psi = {\cal D}^+\Psi = 0$ (or their complex conjugates).
The covariant derivatives here form a set that is closed under (anti)commutation,
as required for consistency of the conditions, which are analogs of
the harmonic analyticity conditions in $N=2, 4D$ supersymmetry \cite{GIKOS}.}

\be
\mbox{either}\ (\mbox{a})\;\; \bar{\cal D}^i \Psi^{(N,M)} = 0
\quad \mbox{or} \  (\mbox{b})\;\;{\cal D}_i \tilde\Psi^{N,M} = 0\,. \label{CHIR}
\ee

\item Equations \p{1d} imply that the $S^2$ covariant derivatives ${\cal D}^+, {\cal D}^+$ form a closed
subset with $\bar{\cal D}^i$ or ${\cal D}_i$ and so preserve chirality.
In other words, they yield some chiral (anti-chiral) superfield when acting on $\Psi^{(N,M)}$
or $\tilde{\Psi}^{(N,M)}$, as defined in \p{CHIR}. Since these derivatives carry non-zero $U(1)$
charges, the charges are shifted from $(N,M)$ to $(N-1, M+1/2)$ for ${\cal D}_+\Psi^{(N,M)}$
and from $(N,M)$ to $(N+1, M-1/2)$ for ${\cal D}^+\Psi^{(N,M)}$. In what
follows we restrict our attention to the chiral superfields.

\item One can consistently require chiral superfields to be covariantly  holomorphic:
\be
\mbox{either}\ \mbox{(a)} \;\;{\cal D}_+\Psi^{(N,M)} = 0 \quad \mbox{or} \  \mbox{(b)}\;\;
{\cal D}^+\Psi^{(N,M)} = 0\,. \label{holom}
\ee
However, a chiral superfield satisfying condition (a)  is zero if $N>0$ and one satisfying
condition (b) is zero if $N<0$.  For chiral superfields with $N=0$ one can
impose {\it both} conditions \p{holom}, thus fully suppressing their $z, \bar z$ dependence.

\item Equations \p{1b}, \p{1c} imply that for $M=0$ or $M=-N$ the covariant derivatives
${\cal D}_2$ and ${\cal D}^+$ or ${\cal D}_1$ and ${\cal D}_+$ together with $\bar{\cal D}^i$ form
a set that is closed under (anti)commutation. Hence the chiral superfields with $M=0$ or $M=-N$
can be subjected to the more stringent set of constraints
\be
{\cal D}_2\Psi^{(N,0)} = \bar{\cal D}^i \Psi^{(N,0)} = 0\,,
\quad {\cal D}^+\Psi^{(N,0)} = 0\,. \label{2a}
\ee
Alternatively, one can impose the constraints
\be
{\cal D}_1\Psi^{(N,-N)} = \bar{\cal D}^i \Psi^{(N,-N)} = 0\,,
\quad  {\cal D}_+\Psi^{(N,-N)} = 0\,. \label{2b}
\ee
Thus chiral superfields can be made `covariantly independent' of one more Grassmann coordinate,
provided they are simultaneously assumed to be holomorphic or antiholomorphic, for $N\geq 0$ or
$N\leq 0$, respectively. In what follows we shall deal with $N\geq 0$, thus specializing to the case (\ref{2a}).

\end{itemize}

For every set of conditions that may be imposed consistently on a superfield  there is a corresponding
invariant subset of the original coordinate set
\be
{\cal Z} = (z, \bar z, \xi^1, \xi^2, \bar\xi_1, \bar\xi_2)\,. \label{ZZ}
\ee
As already mentioned, $(z, \xi^i)$ is one such invariant subset, but there are others. For example,
consider the new non-self-conjugate `chiral' parametrization of SF:
\be
\tilde{\cal Z} = (z, \bar{z}_{sh}, \xi^1, \xi^2, \bar\xi_1, \bar\xi_2)\, \label{ZZZ}
\ee
where
\be\label{shifted}
\bar{z}_{sh} = \bar z - (\xi^2 + \bar z\xi^1)(\bar\xi_1 - \bar z \bar\xi_2)\,.
\ee
One can check that
\be
\delta \bar z_{sh} = \bar a + a\bar{z}_{sh}^2 +
(\bar\epsilon_1 - \bar z_{sh}\bar\epsilon_2)(\xi^2 + \bar z_{sh} \xi^1)\,, \label{Trsh1}
\ee
so the `chiral' subspace
\be
\zeta_L = (z, \bar z_{sh}, \xi^i) \label{zeta}
\ee
is closed under the action of $SU(2|1)$.

To see that the $SU(2|1)$ invariance of the chiral subspace of superspace
is related to the existence of chiral superfields, we set
\be\label{Rescale}
\Psi^{(N,M)} = K_1^{M}K_2^{-N}\Phi^{(N,M)},
\ee
and observe that
\bea
&& \bar{\cal D}^1\Psi^{(N,M)} = K_1^{M-1/2}K_2^{-N-1/2}\bar\nabla^1\Phi^{(N,M)}\,, \nn
&&
\bar{\cal D}^2\Psi^{(N,M)} = K_1^{M}K_2^{-N+1/2}\bar\nabla^2\Phi^{(N,M)}\,, \nn
&& {\cal D}^+ \Psi^{(N,M)} = K_1^{M-1/2}K_2^{-(N+1)}\nabla^{(N)ch}_{\bar z}\Phi^{(N,M)}
\eea
where
\be\label{nablas}
\bar\nabla^1 = K_1 K_2 \bar D^1\,, \quad \bar\nabla^2 = K_2^{-\frac{1}{2}} \bar D^2\,,
\quad  \nabla^{(N)ch}_{\bar z} = K_1^{\frac{1}{2}}K_2 D_- = K_1 K_2^2 \partial_{\bar z}\,.
\ee
{}From the transformation law \p{TranSf}, and the transformations
\bea
\delta K_1 &=& (\bar\epsilon\cdot \xi + \bar\xi\cdot \epsilon)K_1\,, \nn
\delta K_2 &= & \left[a\bar z + \bar a z -\bar\epsilon_1(\xi^1 -z\xi^2)
- (\bar\xi_1 - \bar z \bar\xi_2)\epsilon^1 \right]K_2 ,\label{tranK}
\eea
one can show that
\be
\delta\Phi^{(N,M)} = 2\left\{N[\bar a z - \bar\epsilon_1(\xi^1 - z \xi^2)] -M(\bar\epsilon\cdot \xi)
\right\}\Phi^{(N,M)}\,.\label{TransPhi}
\ee
The next step is to observe that, in the basis \p{ZZZ},
\bea
&& \bar\nabla^1 = -K_1\left\{ \left[ 1 - \bar\xi_1(\xi^1 - z \xi^2)\right]\partial_{\bar{\xi}_1} -
\left[ z + \bar\xi_2(\xi^1 - z \xi^2)\right]\partial_{\bar{\xi}_{2}} \right\}, \nn
&& \bar\nabla^2 = - K_2^{-1}\left(\partial_{\bar{\xi}_{2}} + \bar z \partial_{\bar{\xi}_{1}} \right)\,,
\eea
while $\nabla_{\bar z} \sim \partial_{\bar{z}_{sh}}$. Thus, in the new basis the chirality
constraint (\ref{CHIR}a) becomes
\be\label{Chirphi}
\partial_{\bar{\xi}_1} \Phi^{(N,M)} = \partial_{\bar{\xi}_2}\Phi^{(N,M)} = 0\,\quad \Rightarrow \quad
\Phi^{(N,M)} = \Phi^{(N,M)}(\zeta_L)\,.
\ee

The chiral basis also simplifies the covariant analyticity condition
${\cal D}^+\Psi = 0$ that can be imposed on a chiral superfield $\Psi^{(N,M)}$ because it implies
\be
\nabla_{\bar z}\Phi^{(N,M)}(\zeta_L) = 0\quad \Rightarrow \quad
\Phi^{(N,M)} = \Phi^{(N,M)}(z, \xi^i)\,.\label{Holomor}
\ee
One might describe this state of affairs by saying that the operator ${\cal D}^+$ is `short'
in the chiral  basis, in which case it is worth noting, in contrast, that ${\cal D}_+$
does not share this property  because
\bea
&&{\cal D}_+\Psi^{(N,M)} = K_1^{M+1/2}K_2^{-(N-1)}\nabla_z^{(N)ch}\Phi^{(N,M)}\,,\nn
&& \nabla^{(N)ch}_z = \partial_z -2iN {\cal A}_z = \partial_z -2N \partial_z\log K_2 =
\partial_z -2N\frac{\bar{z}_{sh}}{1 + z\bar{z}_{sh}}\,. \label{chnabla}
\eea

The possibility of imposing the further conditions \p{2a} or \p{2b} on chiral superfields reflects
the existence of the two invariant subspaces
\be
\mbox{(a)} \;\;(z,\; \xi^1 - z\xi^2) \quad \mbox{and} \quad \mbox{(b)}\;\;
(\bar{z}_{sh},\; \xi^2 + \bar{z}_{sh}\xi^1)\,.
\label{twossp}
\ee
The $SU(2|1)$  invariance can be established by noting that
\bea
\delta (\xi^1 - z\xi^2) &=& \epsilon^1 - z\epsilon^2 +\bar a z (\xi^1 - z\xi^2)\,,\nn
\delta (\xi^2 + \bar z_{sh}\xi^1) &=& \epsilon^2 + \bar{z}_{sh}\epsilon^1 +
a \bar{z}_{sh} (\xi^2 + \bar{z}_{sh}\xi^1)\,,\label{Trsh2}
\eea
and using the transformations of $z$ and $\bar{z}_{sh}$ given in \p{transf} and \p{Trsh1}.
These subspaces can be identified with $CP^{(1|1)}$, which is a (holomorphic) supersphere \cite{IMT},
and its dual, the anti-holomorphic supersphere.

Finally, let us see how the more stringent set of conditions \p{2a} with $M=0$ is transformed into
a constraint on the $\xi^i$ dependence of $\Phi^{(N,M)}(z, \xi^i)$ defined in \p{Holomor}.
At $M=0$ the connection term drops out from ${\cal D}_2$, and we have
\bea
&& {\cal D}_2\Psi^{(N,0)} = {D}_2\Psi^{(N,0)} = K_2^{-N-1/2} \nabla_2\Phi^{(N,0)}\,, \nn
&& \nabla_2 = z\partial_{\xi^1} + \partial_{\xi^2}\,.
\eea
Thus the extra condition in \p{2a} is reduced to
\be
\nabla_2\Phi^{(N,0)}(z, \xi^i) = 0 \quad \Rightarrow  \quad
\Phi^{(N,0)} = \Phi^{(N,0)}(z, \xi^1 - z\xi^2)\,. \label{2ared}
\ee
%The same is true as regards the additional Grassmann analyticity condition with ${\cal D}_1$
%in \p{2b}.
%Both the latter and the covariant anti-analyticity condition can also be phrased
%given a form of the explicit Grassmann and bosonic Cauchi-Riemann conditions.
%The corresponding superfield $\Psi^{(N,-N)}$ is expressed through a function
%living on the subspace (\ref{twossp}b).

%[{\bf We should include the two $U(1)$ transformations here}.]
\setcounter{equation}{0}
\section {Super-flag quantum mechanics}
\label{sec.QM}

We now aim to formulate the dynamics of a particle on SF. We shall see that this leads naturally to
superfields of the type described above. We begin by re-interpreting the  1-forms
$(E^A, {\cal A},{\cal B})$ as the corresponding 1-forms induced on the particle's worldline. Thus, we now have
\be
E^A = dt\,  \omega^A, \qquad \omega^A \equiv \dot z E_z^A + \dot\xi^i
E_i{}^A
\ee
and
\bea
{\cal A} = dt\, A , && A\equiv \left[\dot z {\cal A}_z + \dot \xi^i {\cal A}_i\right]
+ c.c.,\nn
{\cal B} = dt\,  B && B\equiv \, \dot\xi^i {\cal B}_i + c.c.\, .
\eea
Note the absence of a $\dot z$-term in $B$. The coefficients $\omega^A=(\omega^+,\omega^1,\omega^2)$
can be used to construct
$SU(2|1)$-invariant kinetic terms, but a term quadratic in
$\omega^i$ would be a `higher-derivative' term that would effectively
double the number of fermion variables. Fortunately, there is no need to
include such a term; we may construct an $SU(2|1)$ invariant kinetic
term from $\omega^+$ alone. Although it also contains terms with
derivatives of the `fermi' variables $\xi^i$, these occur only in nilpotent
`fermion'-bilinear terms. Specifically,
\be
\omega^+ = \dot z \omega + \dot\xi^i \omega_i
\ee
where
\bea
\omega &=& K_1^{-{1\over2}}K_2^{-1}\,,\nn
\omega_1 &=& -K_1^{-{3\over2}}K_2^{-1}\left(\bar\xi_2 +
z\bar\xi_1\right), \nn
\omega_2 &=&  K_1^{-{3\over2}}K_2^{-1} z \left(\bar\xi_2 +
z\bar\xi_1\right).
\eea
Note that $\omega$ happens to be real, although all other coefficients
are complex. We will see soon that the presence of the $\dot\xi^i$ terms
in $\omega^+$ is innocuous. In addition to the kinetic term, there are two possible WZ terms that we may construct from $A$ and $B$. We record here that
\bea
{\cal A}_z &=& -iK_2^{-1}\left[\bar z -\xi^2\left(\bar\xi_1
-\bar z\bar\xi_2\right)\right], \nn
{\cal A}_1 &=& -iK_2^{-1}\left(\bar\xi_1 -\bar z\xi_2\right), \nn
{\cal A}_2 &=&  iK_2^{-1}z\left(\bar\xi_1 -\bar z\xi_2\right),
\eea
and
\be
{\cal B}_i =-iK_1^{-1}\bar\xi_i.
\ee

These considerations lead us to consider the Lagrangian
\be
L = |\omega^+|^2 + NA + MB \label{Lagr}
\ee
where $N$ and $M$ are two constants. Let $(p,\pi_1,\pi_2)$ be the variables canonically
conjugate to $(z,\xi^1,\xi^2)$. An alternative, phase-space, Lagrangian is then
\be
L= \left\{ \left[\dot z p +i \dot\xi^i \pi_i  +
\lambda^i\varphi_i\right] +
c.c.\right\} - H
\ee
where $H$ is the Hamiltonian
\be\label{hamiltonian}
H= \omega^{-2}|p- N{\cal A}_z|^2,
\ee
and $\lambda^i$ $(i=1,2)$ is a pair of complex Grassmann-odd Lagrange
multipliers for the complex Grassmann-odd constraints $\varphi_i\approx 0$,
where
\be
\varphi_i = \pi_i +i\omega^{-1}\omega_i\left(p- N{\cal A}_z\right)  +i N{\cal A}_i +
i M{\cal B}_i\,. \label{Constr1}
\ee
Taken together with their complex conjugates, these constraints are second
class, in Dirac's terminology. However, they are first class if viewed as
two holomorphic constraints. Following the `Gupta-Bleuler' method of
dealing with complex second class constraints, as recently explained in
the context of CSQM models in \cite{IMPT,IMT}, we may view the constraints
$\varphi_i\approx 0$ as gauge-fixing conditions for gauge invariances
generated by their complex conjugates $\bar\varphi^i$. Stepping back to
the gauge-unfixed theory, we may then quantize initially {\it without
constraint} by setting
\be\label{correspond1}
p =  -i{\partial\over\partial z} \, ,\qquad \bar p =
-i{\partial\over\partial \bar z}
\ee
and
\be\label{correspond2}
\pi_i = {\partial\over\partial \xi^i}\, ,\qquad
\bar\pi^i = {\partial\over\partial \bar\xi_i}.
\ee
The constraint functions $\bar\varphi^i$ then become the complex operators
\be
\hat {\bar\varphi}^i = {\partial\over\partial \bar\xi_i} - \omega^{-1}
\bar\omega^i \left[ {\partial\over\partial \bar z} +
N{\partial \log K_2\over\partial \bar z}\right] + N{\partial \log K_2
\over
\partial \bar\xi_i} -M  {\partial \log K_1\over\partial \bar\xi_i}\,.
\ee

To take the constraints into account it is now sufficient to impose the physical state conditions
\be
\hat {\bar\varphi}^i |\Psi\rangle =0 \qquad (i=1,2).
\ee
We will solve this constraint in two steps. The first step, suggested by \p{Rescale}, is to set
\be\label{reduced}
\Psi = K_1^{M} K_2^{-N} \Phi
\ee
for `reduced' wavefunction $\Phi$, for which the physical state conditions are
\be
\left({\partial \over \partial \bar\xi_i} - \omega^{-1}\bar\omega^i
{\partial \over \partial \bar z}\right)\Phi =0 \qquad (i=1,2).
\ee
These are equivalent to the two conditions
%[{\bf I seem to differ by a
%sign
%in the second one, as compared to Luca}.] [{\bf I suppose that you changed
%the definition of the fermion momenta from that used in the odd coset paper!!!!}]
\be
\left[{\partial \over \partial \bar\xi_2} + \bar z \left({\partial \over
\partial \bar\xi_1}\right)\right]\Phi=0\, , \qquad
\left[{\partial \over \partial \bar\xi_1} - K_1^{-1} \left(\xi^2 + \bar
z\xi^1\right) {\partial \over \partial \bar z}\right]\Phi =0\,.
\ee
These conditions are equivalent to the chirality conditions
%[{\bf Actually, they aren't, either
%for
%these or Luca's original constraints. After the typo corrected before it should!!!! Some errors somewhere.}]
\be
\bar{D}^i\Phi = 0 \quad \mbox{or} \quad \bar \nabla^i\Phi=0\,, \qquad (i=1,2)
\ee
where $\bar\nabla^i$ were defined in \p{nablas}. In other words, the reduced wavefunction is `chiral',
with $N$ and $M$ being two $U(1)$ charges. The general solution of such chirality constraints
was given in \p{Chirphi}:
\be\label{shifted1}
\Phi = \tilde \Phi\left(z,\bar z_{sh},\xi^1,\xi^2\right)
\ee
where $\bar z_{sh}$ is the `shifted' coordinate defined in \p{shifted}.
%\be
%\bar z_{sh} = \bar z - K_1^{-1}\left(\xi^2 +\bar
%z\xi^1\right)\left(\bar\xi_1 -\bar z\bar\xi_2\right).
%\ee
The function $\tilde \Phi$ can be expanded in a terminating Taylor
series
in $\xi^1,\xi^2$. Each of the four independent coefficient functions is
determined by a single function on $S^2$, two of which are Grassmann-odd
and two Grassmann-even.

The $SU(2|1)$ invariance of our model implies the existence of corresponding Noether charges.
In particular, there exist Grassmann-odd Noether charges which, upon quantization become the
operators
\bea
\hat S_1 &=& {\partial\over\partial\xi^1} + M\bar\xi_1 -
\bar\xi_1\left(\bar\xi \cdot {\partial \over \partial\bar \xi}\right) +
\left({\bar \xi}_1 - {\bar z}{\bar \xi}_2\right){\bar z}{\partial_{\bar
z}}
+{N} \left({\bar \xi}_1 - {\bar z}{\bar \xi}_2\right), \nn
\hat S_2 &=& {\partial\over\partial\xi^2} + M\bar\xi_2 -
\bar\xi_2\left(\bar\xi \cdot {\partial \over \partial\bar \xi}\right) +
\left({\bar \xi}_1 - {\bar z}{\bar \xi}_2\right)
{\partial_{\bar z}}  , \nn
\hat {\bar S}{}^1 &=& {\partial\over\partial\bar\xi_1} + M\xi^1 +
\xi^1{\left(\xi\cdot {\partial \over \partial \xi}\right)} -
\left({\xi}^1 - {z}{ \xi}^2\right)z{\partial_{ z}} +{N}
\left({\xi}^1 -
{z}{ \xi}^2\right),\nn
\hat {\bar S}{}^2 &=& {\partial\over\partial\bar\xi_2} +
M\xi^2 +
\xi^2{\left(\xi\cdot {\partial \over \partial \xi}\right)} -
\left({\xi}^1 - {z}{ \xi}^2\right){\partial_{z}}, \label{Sgen}
\eea
These operators weakly anticommute with the constraints \p{Constr1}.
Their non-zero anticommutation relations are
\bea
&&\{ \hat{S}_i, \hat{S_k} \} = \{\hat{\bar S}{}^i, \hat{\bar S}{}^k \} = 0\,, \nn
&& \{\hat{S}_1, \hat{\bar S}{}^1\} = J_3 + B\,, \quad
\{\hat{S}_2, \hat{\bar S}{}^2\} = B\,, \nn
&& \{\hat{S}_1, \hat{\bar S}{}^2\} = -J_+\,, \quad
\{\hat{S}_2, \hat{\bar S}{}^1\} = J_-\,, \label{Alg}
\eea
where
\bea
&& B = z\partial_z - \bar z \partial_{\bar z} + \left(\xi^1\frac{\partial}{\partial \xi^1}
-\bar\xi_1\frac{\partial}{\partial \bar\xi_1}\right) + 2M\,,\nn
&&\quad J_3 = \left(\xi^2\frac{\partial}{\partial \xi^2}
- \xi^1\frac{\partial}{\partial \xi^1}\right) - \left(\bar\xi_2\frac{\partial}{\partial \bar\xi_2}
- \bar\xi_1\frac{\partial}{\partial \bar\xi_1}\right)
-2(z\partial_z - \bar z\partial_{\bar z}) + 2N\,, \nn
&& J_+ = \partial_z + (\bar z)^2\partial_{\bar z} +\left(\xi_2\frac{\partial}{\partial \xi_1}
- \bar\xi_1\frac{\partial}{\partial \bar\xi_2}\right) + \bar z N\,, \nn
&& J_- = \partial_{\bar z} + z^2\partial_{z} -\left(\xi_1\frac{\partial}{\partial \xi_2}
- \bar\xi_2\frac{\partial}{\partial \bar\xi_1}\right) - z N\,. \label{anticomm}
\eea
%[{\bf N, M are redefined, as in Sect 1!!!}].
Acting on the coordinates $Z=(z, \bar z, \xi^i, \bar \xi_i)$ these operators generate the transformations
\p{transf}, and hence the transformations \p{TranSf}  of  $\Psi(Z)$ for a superfield  with $U(1)$ charges
$M$ and $N$.

\setcounter{equation}{0}
\section{Super-Landau levels}

We have just seen that the wavefunction of a particle on SF is a chiral superfield.
Of course, a general wavefunction will also be time-dependent  but
it can be expanded on a basis of stationary states with time-dependent coefficients that
depend on the energy eigenvalues. These stationary states are time-independent chiral superfields,
and  our next task is to determine the energy eigenvalues and also the type of chiral superfield
at each level. As we shall see, the ground state chiral superfield is one for which the
reduced wavefunction is analytic.

Using the correspondence (\ref{correspond1}) we have
\be\label{identity}
i(p-N{\cal A}_z) \rightarrow \nabla_z^{(N)} =  \partial_z - i NA_z.
\ee
The quantum Hamiltonian operator $\hat H$ corresponding to the classical Hamiltonian (\ref{hamiltonian})
involves the product of $\nabla_z^{(N)}$ and its
complex conjugate $\nabla_{\bar z}^{(N)}$. The product is ambiguous because, from  \p{1A}, \p{Dnabla}, \p{commnabla},
\be\label{comm}
\left[{\cal D}_+, {\cal D}^+\right] = K_2^2K_1\left[\nabla_z^{(N)},\nabla_{\bar z}^{(N)}\right] =2N\,.
\ee
The natural resolution of this ambiguity is to define the quantum Hamiltonian operator to be
\be
\hat H = - {1\over 2} \{{\cal D}_+, {\cal D}^+\} = -{1\over 2}K_2^2K_1 \{ \nabla_z^{(N)} ,\nabla_{\bar z}^{(N)}
\}\,,\label{hat1}
\ee
as this is manifestly positive definite. Equivalently,
\be
\hat H = H_N = -{\cal D}_+{\cal D}^+ + N \equiv -K_2^2K_1 \nabla_z^{(N)} \nabla_{\bar z}^{(N)} + N\,.
\label{hat2}
\ee
One can show that
\be
[\hat H,\hat{\bar\varphi}^i]=0\,,
\ee
so that the Hamiltonian can be consistently restricted to a `reduced
Hamiltonian' operator
\be
\hat H_{red} = K_1^{-M}K_2^{N}\,\hat{H}\,K_1^{M}K_2^{-N} = -\nabla_z^{(N +1)ch}\nabla_{\bar z}^{(N)ch} + N =
-K_2^2K_1 \nabla_z^{(N)ch} \partial_{\bar z} + N\,, \label{hatred}
\ee
which acts on reduced wavefunctions
$\Phi(z,\bar z_{sh},\xi^1,\xi^2)$ (see \p{nablas} and \p{chnabla}
for the definition of $\nabla_{\bar z}^{(N)ch}$
and  $\nabla^{(N)ch}_z$).
Clearly, any holomorphic chiral superfunction $\Phi_0(z, \xi^1,\xi^2)$ is an eigenfunction of $H_{red}$
with eigenvalue $N$. This is the ground state energy, although we postpone the proof of this until
we complete, in the next section,  the characterization of all admissible states, which also involves
a determination of the degeneracies. First we must determine the energy levels,
which we do using Schroedinger's factorization method. This method was recently applied to
another supersymmetric extension of the Landau model for a particle on the 2-sphere \cite{MM}
and what follows here is similar.

Using \p{Dnabla}, we can rewrite \p{hat2} as
\be
H_N = - UV + N
\ee
where
\be
U= K^{\frac{1}{2}} K_2\nabla_z^{(N+1)} , \qquad V= K^{\frac{1}{2}}K_2 \nabla_{\bar z}^{(N)}\, .
\ee
The factorization trick exploits the fact that all non-zero eigenvalues of $H_N$ are also eigenvalues
of the `reverse-order' Hamiltonian
\be
\tilde H_N = - VU +N\,.
\ee
It follows that the first excited state of $\hat H=H_N$ is the ground state of $\tilde H_N$. However, we also have
\bea
\tilde H_N  &= & -K_1K_2 \nabla_{\bar z}^{(N)} K_2 \nabla_z^{(N+1)}  +N
\nn
&=& -K_1K_2^2 \nabla_{\bar z}^{(N+1)} \nabla_z^{(N+1)}  +N \nn
&=& -K_1K_2^2 \nabla_z^{(N+1)} \nabla_{\bar z}^{(N+1)}  + 2(N+1) +N
\eea
where we have used (\ref{comm}) to get to the last line. Thus
\be
\tilde H_N = H_{N+1} + 2N+1\,.
\ee
We know that the ground state energy of $H_{N+1}$ is $N+1$ so we deduce that the first excited state
of $\hat H=H_N$ has energy $3N+2$. The corresponding eigenstate is $\Psi_1 = U\Psi_0={\cal D}_+\Psi_0$,
where $\Psi_0$ is a ground state wavefunction.\footnote{This covariantly chiral wavefunction has the $U(1)$
charges $(M-1/2, N+1)$ and so corresponds
to the ground state of {\it another} system,  with the coefficients $(M-1/2, N+1)$
in the relevant WZ terms.}

By iteration one now deduces that the full set of energy levels are
\be
E= (2\ell + 1)N + \ell(\ell+1)\,, \qquad \ell=0,1,2,\dots \label{levels}
\ee
with the corresponding reduced wavefunctions
\be\Phi^N_{(\ell)} = \nabla_z^{(N+\ell)ch} \dots \nabla_z^{(N+1)ch} \Phi_0\left(z;
\xi^1,\xi^2\right) \qquad (\ell=1,2,\dots)\label{levred}
\ee
where $\Phi_0$ (having the $U(1)$ charges $(M-\ell/2, N+\ell)$) is a ground state reduced wavefunction.

It is worth noting that the easiest way to check that \p{levels}
is indeed the eigenvalue of the original hamiltonian $H_N$ of \p{hat2}, corresponding to the reduced
wavefunction \p{levred}, is to consider the covariantly chiral wavefunction
\be\label{chirll}
\Psi^{(N, M)}_{(\ell)} = K_1^{M}K_2^{-N}\Phi^N_{(\ell)} = ({\cal D}_+)^\ell \Psi_{(0)}^{(N+\ell, M-\ell/2)}\,,
\ee
where here we make explicit  the $U(1)\times U(1)$ charges of $\Psi_{(0)} =K_1^{M-\ell/2}K_2^{-(N+\ell)}\Phi_{0}$.
Acting with $H_N = -{\cal D}_+{\cal D}^+ + N $ on this wavefunction,  taking into account
the $U(1)\times U(1)$ charges of ${\cal D}_+$, the commutation relation \p{1A},  and the covariant
analyticity condition
$$
{\cal D}^+\Psi_{(0)}^{(N+\ell, M-\ell/2)} = 0\,,
$$
it is a matter of simple algebra to show that
\be
H_N \Psi^{(N, M)}_{(\ell)} = [(2\ell+1)N+ \ell(\ell+1)]\Psi^{(N, M)}_{(\ell)}\,.
\ee
Note the absence of any $M$-dependence of these eigenvalues. This makes it appear that
the $U(1)$ charge $M$ does not influence the structure of the Hilbert space. As we shall soon see,
this is far from true.

\setcounter{equation}{0}
\section{Degeneracies}

We now turn to a consideration of the $SU(2|1)$ content of the Hilbert space, which involves consideration
of the Hilbert space norm. The $SU(2|1)$-invariant norm $||\Psi||$ of $\Psi$ is
given by the formula
\be
||\Psi||^2 = \int \!d\mu \, |\Psi|^2 = \int \!d\mu_0\,  K_2^{-2} |\Psi|^2 \label{normpsi}
\ee
where
\be
\int\! d\mu_0 =\int \! d^2 z\, \prod_{i=1}^2
{\partial\over \partial\xi^i}{\partial \over\partial\bar\xi_i}
\ee
and the integral is over all complex $z$ (which covers the sphere except
for the point at infinity that does not contribute to the value of the
integral). This result follows from the fact that
\be
\left(sdet {E_{M}}^{A}\right)\left( sdet {E^{\bar M}}_{\bar A} \right)= K_2^{-2}.
\ee
The $SU(2|1)$ invariance of the measure $d\mu =d\mu_0\, K_2^{-2}$
can be verified using the transformation law \p{tranK} for $K_2$, and
\bea
\delta\left(d\mu_0\right) &=& (\partial_z\delta z - \partial_{\xi^i}\delta \xi^i + \mbox{c.c.})\,d\mu_0 \nn
&=&  2[\bar a z + a \bar z - \bar\epsilon_1(\xi^1-z\xi^2)
- (\bar\xi^1 -\bar z\bar\xi^2)\epsilon^1]\,d\mu_0\,.
\eea
For a physical wavefunction of the form  (\ref{reduced}),  we have
\be\label{normphi}
||\Psi||^2 = \int\! d\mu_0\, K_1^{2M}K_2^{-2(N+1)} |\Phi|^2.
\ee
As we saw in section \ref{sec.superfield}, for chiral $\Psi$ the reduced wavefunction $\Phi$ takes the form
\be\label{shifted2}
\Phi = \tilde \Phi\left(z,\bar z_{sh},\xi^1,\xi^2\right)
\ee
where $\bar z_{sh}$ is the `shifted' coordinate defined in \p{shifted}.

We first evaluate \p{normphi} for the ground state wavefunction $\Psi_0$ for which $\Phi$
is analytic and has the component field expansion
\be
\Phi \left( z, {\xi}^{i} \right)  = A(z)+ \xi^i\psi_i (z)+ \xi^1\xi^2 F(z)\,.  \label{Multipl}
\ee
Using this in (\ref{normphi}) and performing the Berezin integrals, we find that
\bea
||\Psi_0||^2 &=& 2\int{ {dzd{\bar z}\over \left( 1 + z{\bar z}\right)^{2\left(N+1\right)}}
\Bigg[M\left( 2M + 2N +1\right)
|A|^2 + {1\over 2}|F|^2} \nonumber \\
&&+\  M \left(\bar\psi^1\psi_1 + \bar\psi^2\psi_2\right)
+ \frac{N +1}{1 + z {\bar z}}\left( {\bar \psi}^2
+ {\bar z} {\bar \psi}^1\right)\left( \psi_2 + z \psi_1\right)\Bigg]
\,.\label{Norm1}
\eea
For non-zero $M$ we see that the ground-state multiplet contains two complex bosonic
fields $A(z)$ and $F(z)$, as well as an SU(2) doublet of holomorphic Grassmann-odd fields $\psi_i(z)$ ($i=1,2$).
For these to be globally defined on the sphere, their norms should be square-integrable on $S^2$,
i.e. the corresponding pieces of the integral on the right hand side of \p{Norm1}  should converge.
This requires $A(z)$, $F(z)$ and each of the $\psi_i(z)$ to be polynomials of degree $\le 2N$,
which means that they each carry a $(2N+1)$-dimensional, spin $N$, representation  of $SU(2)$.
Actually, as $\psi_i(z)$ form an $SU(2)$ doublet, the Grassmann-odd fields carry
the reducible representation ${\bf [2]}\otimes {\bf [2N+1]} = {\bf [2N +2]}\oplus {\bf [2N]}$
(the last term in \p{Norm1} just involves the irreducible ${\bf [2N+2]}$ part of this $SU(2)$ representation).
Thus we have a total of $4N+2$ bosonic components carried by $A(z)$ and $F(z)$ and $4N+2$
fermionic components carried by $\psi_i(z)$. Their transformation rules under the $U(1)$
charge $B$ are specified by the external overall $B$-charge $M$ and the transformation
properties  \p{u1transf} of the coordinates $(z, \xi^i)$.

{}From this result it is clear that $2N$ must be a positive integer, as expected because
this was true of the bosonic Landau model. It then follows that $M\ge0$ since the norm
of the wavefunction with $\Phi=A(z)$ would otherwise be negative. For $M=0$ this wave-function has zero norm.
In this case the multiplet \p{Multipl} splits into a semi-direct sum of an irreducible multiplet, with fields
\be\label{short}
F(z),\; \chi(z)\,, \quad (\chi\equiv \psi_2+ z \psi_1)\,,
\ee
and a quotient which transforms into this irreducible set. In other words, for $M=0$
we are facing a representation of $SU(2|1)$ that is not-fully reducible.\footnote{This property is
reflected in the structure of the transformation law \p{TransPhi} because the `weight'  piece at $M=0$
becomes a function of the coordinates $(z, \xi^1 - z \xi^2)$, which form a closed set under
the action of $SU(2|1)$; recall that precisely when $M=0$ one can consistently impose
on the holomorphic chiral superfield the additional Grassmann analyticity conditions \p{2a},
which forces it to `live' on this smaller space. In terms of the component fields,
this additional covariant constraint amounts to setting to zero the irreducible set $\left(F(z),\chi(z)\right)$,
after which the quotient becomes the degenerate irreducible ${\bf [2N+1]}\oplus {\bf [2N]}$,
`superspin' $N$, multiplet. Though the norm \p{Norm1} is vanishing for the latter,
one can presumably define for it an alternative $SU(2|1)$ invariant norm which
is positive-definite (see \cite{IMPT}). We shall not dwell further on this  possibility
since it is unclear how to incorporate the conditions \p{2a}, \p{2ared}
into our analyticity quantization method. Indeed, they inevitably require ${\cal D}^+\approx 0$,
which does not arise as a constraint
within the hamiltonian formalism in our model, although it does in the Lowest Landau Level
limit in which the kinetic term of $z, \bar z$ is suppressed in \p{Lagr}.
So, this possibility would be of
interest to study in the framework of Chern-Simons Quantum Mechanics on $SU(2|1)$.}
Normalizability implies that  $F(z)$ is a (Grassmann-even) polynomial of degree $\le 2N$,
and that $\chi(z) = \psi_2(z) + z\psi_1(z)$ is a (Grassmann-odd) polynomial of degree $\le 2N+1$.
The $SU(2)$ content in this case is therefore ${\bf [2N+1]}\oplus {\bf [2N+2]}$ and these
combine to yield the degenerate,  `superspin' $\left(N+{1\over2}\right)$, irrep of $SU(2|1)$,
of the type carried by a LLL particle on  the supersphere \cite{IMT}.

Now we turn to the case of a general chiral superfield wavefunction, for which
the reduced wavefunction depends both on $z$ and on
\be
\bar z_{sh} = \bar z - v, \qquad v\equiv \left(\xi^2 +\bar z\xi^1\right)\left(\bar\xi_1 -\bar z\bar\xi_2\right).
\ee
As $H$ is nilpotent,
\be
\Phi = \tilde\Phi \left( z, {\bar z}, {\xi}^{i} \right)  -
v\partial_{\bar z} \tilde\Phi \left( z, {\bar z}, {\xi}^{i} \right).
\ee
Using the component field expansion
\be
\tilde\Phi \left( z, {\bar z}, {\xi}^{i} \right) =
\tilde A(z,\bar z)+ \xi^i\tilde \psi_i (z,\bar z)+ \xi^1\xi^2 \tilde F(z,\bar z)\label{ComP}
\ee
we find that
\bea\label{Norm3}
||\Psi||^2 &=&  ||\Psi||_0^2  - \int \!{ {dzd{\bar z}\over \left( 1 + z{\bar z}\right)^{2\left(N+1\right)}}
\bigg[\left( 1 + z{\bar z}\right)^2 |\partial_{\bar z}\tilde{A}|^2}\nn
&& +\, \left\{ \left( \tilde{\bar \psi}{}^2 + {\bar z} \tilde{\bar \psi}{}^1\right)
\left( \partial_{\bar z}\tilde{\psi}_1 -{\bar z} \partial_{\bar z}\tilde{\psi}_2\right) + h.c.  \right\}\bigg]
\eea
where $||\Psi||_0$ is the norm as it would be if we were dealing with the ground state
(that is, the same as the ground state norm in (\ref{Norm1}) but with non-holomorphic component fields
defined in \p{ComP}).

Notice the relative minus sign in (\ref{Norm3}). Let us see what effect this has
on the first excited state wavefunction $\Psi_1$,  for which
\be
\tilde\Phi = {\nabla_{z}}^{\left( N + 1\right)ch}\Phi\left( z,\xi^{i}\right) \,.
\ee
In terms of the holomorphic component fields of the ground state reduced wavefunction
the component fields of the first excited state are
\bea
A^{(1)} & = &\left({\partial_{z}} - {{2\left(N+1\right){\bar z}}\over{1 + z{\bar z}}}\right)A\left(z\right),\nn
\psi_{i}^{(1)}& = &\left({\partial_{z}} - {{2\left(N+1\right){\bar z}}\over{1 + z{\bar z}}}\right)
\psi_{i}\left(z\right),\nn
F^{(1)} & = &\left({\partial_{z}} - {{2\left(N+1\right){\bar z}}\over{1 + z{\bar z}}}\right)F\left(z\right).
\eea
The derivatives with respect to $\bar z$ appearing in (\ref{Norm3}) are now trivially computed.
After integrating by parts with respect to both $\partial_z$ and $\partial_{\bar z}$ we arrive
at the surprising result that the norm $||\Psi_1||^2$ coincides, up to a factor,  with $||\Psi_1||^2_0$,
which has the same form as \p{Norm1} but with $M\rightarrow M-1/2\,, \; N \rightarrow N+1$.
Thus the norm of the $\Psi_1$
is positive  iff $M\geq 1/2$. Clearly, the norm of $\Psi_0$ is also positive under the same condition on $M$.

This result has the following generalization. The  $\ell$th Landau level wavefunction has positive norm
provided that
\be
M \geq \ell/2\,, \label{bound}
\ee
It  follows that for  fixed $M$ the physical Hilbert space is spanned by the states with
\be
0 \leq \ell \leq 2[M]\,. \label{interval}
\ee
In other words, the number of Landau levels  is {\it finite} in this model, a striking contrast
with the bosonic problem for which the number of levels is infinite.

To prove this general result it is convenient to work with covariantly chiral wavefunctions,
and we begin with the first level for which the corresponding covariantly chiral wavefunction is
\be \label{1stLL}
\Psi_{(1)}^{(N,M)} = {\cal D}_+\Psi_{(0)}^{(N+1, M-1/2)}\,, \quad {\cal D}^+\Psi_{(0)}^{(N+1, M-1/2)} = 0\,.
\ee
Substituting this into the norm as given in \p{normpsi},  and integrating by parts
with respect to $\partial_{\bar z}$, it is easy to bring the norm into the form
\be
||\Psi_{\left(1\right)}^{(N,M)}||^2 =
- \int d\mu_0\,  K_2^{-2}\ \overline{\Psi}_{(1)}{\,}^{(-N-1,-M +1/2)}{\cal D}^+{\cal D}_+
\Psi_{{\left(0\right)}}^{(N+1, M-1/2)}\,.
\ee
Pulling  ${\cal D}^+$ out to the right, using the commutation relation \p{1A}, taking into account
the $U(1)\times U(1)$ charges and using the analyticity condition in \p{1stLL}, one deduces that
\be
||\Psi_{\left(1\right)}^{(N,M)}||^2 = 2(N+1)\int d\mu_0\,  K_2^{-2} \left|\Psi_{\left(0\right)}^{(N+1, M-1/2)}\right|^2\,.
\ee
This differs from the ground-state norm by the factor $2(N+1)$ and the shift
$(N,M) \rightarrow (N+1, M-1/2)$. This is just what we found before by direct evaluation in components.

The same method applied to the norm of the $\ell$-level wavefunction \p{chirll}, yields the result
\be
||\Psi_{\left(\ell\right)}||^2 = \ell !\, \frac{(2N +\ell+1)!}{(2N +1)!}\int d\mu_0\,
K_2^{-2} \left|\Psi_{(0)}^{(N+\ell, M-\ell/2)}\right|^2\,.
\ee
Hence, up to the positive factor, this norm is given by the expression \p{Norm1} with $M \rightarrow M -\ell/2$ and
$N \rightarrow N+\ell$. From this, the bound \p{bound} and the restriction \p{interval} follow. In terms of
the eigenvalue $F$ of the $U(1)$ operator $\hat F$ commuting with $SU(2)$ and defined in \p{hatF}, the restriction
\p{bound} is
\be
\ell \leq {1\over 2}F - N\,. \label{bound1}
\ee

Finally, we note that in the sector of all admissible states it is easy to show that $N$ in \p{hat2} indeed
provides the lowest energy. One sandwiches the first term in \p{hat2} between arbitrary physical
states and finds that this average is always $\geq 0\,$.

\section{Concluding remarks}

We have presented an $SU(2|1)$ invariant extension of
the $SU(2)$-invariant Landau model for a particle on $S^2$,
depending on $U(1)$ charges $2N$ and $2M$. In our case, the particle moves
on the superflag manifold $SU(2|1)/[U(1)\times U(1)]$, which is a supermanifold
of complex dimension $(1|2)$ having $S^2$ as its body. As was to be expected,
the Hilbert superspace of each Landau level carries an irreducible representation of $SU(2|1)$,
which depends on $N$, but, surprisingly, the number of admissible levels is finite, being determined by $M$.

Also notable is the fact that  if $2M$ is an integer then the Hilbert superspace of the
last admissible level (at  $\ell = 2M$)  carries a {\it degenerate} representation of $SU(2|1)$
corresponding to a wavefunction in a short  supermultiplet. In particular, if $M=0$ then only
the lowest Landau level is admissible, and  we effectively have a LLL model for a particle
on the superflag, which defines a fuzzy superflag. One might have expected the $N\rightarrow\infty$
limit to yield a classical superflag but the $SU(2|1)$ content of  its LLL Hilbert space
coincides with the $SU(2|1)$ content of an LLL model for a particle on the supersphere,
and this yields the classical supersphere in the large representation limit  \cite{IMT}).

Another notable feature, shared with the bosonic model, is that wavefunctions
of any admissible Landau level  for fixed $N$ and $M$ are expressed in terms of
the ground state functions of a similar model, but with other values of these $U(1)$ charges.
Since the ground states correspond to lowest Landau levels, and hence to some topological
Chern-Simons mechanics, we deduce that the Hilbert space of  the full Landau problem
is the sum of Hilbert spaces for a set of inequivalent  LLL models for a particle
on $SU(2|1)/[U(1)\times U(1)]$.

As some avenues for further study, let us mention that  we are not aware of any comparable
analysis of  the bosonic  $SU(3)/[U(1)\times U(1)]$ `Landau' model.  One might also wish for a formulation  that is manifestly independent of the parametrization of the coset (super)space,
as can be achieved via the introduction of harmonic variables  \cite{GIKOS}.

\section*{Acknowledgements}
We thank Anatoly Pashnev for discussions and collaboration at the early 
stage of this study. We also acknowledge useful discussions with  
M. Berkooz, S. Krivonos and L. Susskind. The work of E.I. was partially 
supported by the INTAS grant No 00-00254, RFBR-DFG grant No 02-02-04002, 
grant DFG No 436 RUS 113/669 and RFBR grant No 03-02-17440. He is grateful 
to the Physical Department of the University of Miami for the warm 
hospitality extended to him during the course of this work. L.M. thanks 
the Bogoliubov Laboratory of Theoretical Physics at JINR, the Physics 
Department of the Weizmann Institute, the General Relativity Group of DAMTP, 
and the Theory Group of the Stanford University, for the hospitality 
and partial financial support.  Some of the results presented here have 
also been part of a talk of L.M. at the {\sl Workshop on Branes and 
Generalized Dynamics} (Argonne, October 20-24, 2003).  This work was 
supported in part by the National Science Foundation under the 
grant PHY-9870101.

\end{document}